# Impurity diffusion in highly-ordered intermetallic compounds studied by nuclear quadrupole interactions

Gary S. Collins [a *], Qiaoming Wang [b] and John P. Bevington [c]

Department of Physics and Astronomy, Washington State University, Pullman, WA 99164 USA

[a] collins@wsu.edu, [b] qmwangustc@gmail.com, [c] jbev85@gmail.com

**Abstract.** Diffusion of impurity atoms depends on the sublattices occupied, active diffusion mechanisms, and jump frequencies to neighboring sites. The method of perturbed angular correlation of gamma rays (PAC) has been applied over the past decade to study impurity diffusion through measurement of nuclear quadrupole interactions (NQI) at nuclei of $^{111}$In/Cd probe atoms. Extensive measurements have been made on highly-ordered compounds having the L1$_2$ crystal structure, including In$_3$R, Sn$_3$R, Ga$_3$R, Al$_3$R and Pd$_3$R phases (R= rare-earth element). Measurements in thermal equilibrium at high temperature served to determine lattice locations of $^{111}$In parent probe-atoms, through characteristic NQIs, and to measure diffusional jump-frequencies of $^{111}$Cd daughter probe-atoms, through relaxation of the NQI. This paper summarizes results of the jump-frequency measurements and relates them to the conventional diffusivity that can be determined, for example, from penetration profiles of tracer species. In spite of chemical similarities of the series of rare-earth phases studied, remarkably large variations in jump frequencies have been observed especially along series of In$_3$R and Pd$_3$R phases. Most phases appear as "line compounds" in binary phase diagrams, but large differences in site-preferences and jump-frequencies were observed for samples prepared to have the opposing limiting phase boundary compositions. Comparisons of jump-frequencies measured at opposing boundary compositions can give insight into the predominant microscopic diffusional mechanisms of the impurity. A change in diffusion mechanism was proposed in 2009 to explain jump-frequency systematics for In$_3$R phases. An alternative explanation is proposed in the present paper based on site-preferences of $^{111}$Cd daughter probes newly observed along the parallel Pd$_3$R series. The diffusivity can be expressed as the product of a jump-frequency such as measured in these studies and a correlation factor for diffusion that depends on the diffusion mechanism. The correlation factor can be modeled for the L1$_2$ structure and diffusion sublattice of interest using a five-frequency model originally proposed for metals. Although the correlation factor is an essential parameter for the diffusion of impurities, it has never been measured. It is suggested that values of the correlation factor can be determined feasibly by combining results of jump-frequency measurements such as the present ones with diffusivity measurements made for the same host-impurity systems.

## Introduction

There is considerable interest in diffusion in compounds. In intermetallic compounds, screening by conduction electrons reduces the effective charges of the ion cores of constituent host atoms. As a result, antisite atoms have lower formation enthalpies than they would in ionic compounds and are commonly observed defects. Consider a binary compound that has one sublattice for each element. When host atoms have comparable sizes and the crystal structure is close-packed, as in the present case, the intrinsic point defects of importance are vacancies and antisite atoms on the two sublattices. Phases studied in this paper appear mostly as "line compounds" in binary phase diagrams, meaning that widths of the phase field are of the order of 1 at.% or less and difficult to measure with precision. In a general way, this can be attributed to high partial formation enthalpies of the intrinsic point defects, leading also to a high degree of crystalline order of the phases.

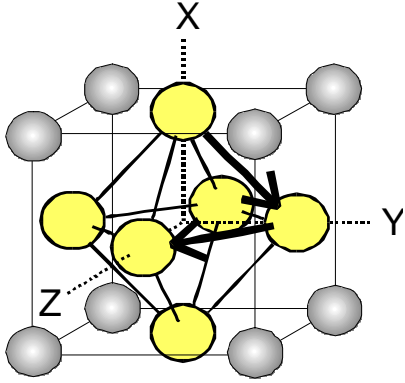

*Fig. 1*. The L1$_2$ crystal structure, with generic formula A$_3$B, with A atoms (face-center sites) equal to an sp-element or Pd, and B atoms (corner sites) to a rare-earth element. Arrows represent two successive, diffusive jumps of an atom on the A-sublattice, caused by rapid passage of vacancies on the A-sublattice (not shown). Jumps lead to decoherence of the nuclear spin precessions that are detected using the PAC technique and that can be fitted to obtain accurate values of the mean jump-frequencies.

**Self-diffusion** in intermetallic compounds has been extensively studied using a wide variety of experimental methods [1, 2, 3]. Out of these studies has come a general understanding of diffusion mechanisms in common and technologically important crystal structures. The focus of the present paper is on diffusion in compounds having the Cu$_3$Au, or L1$_2$, crystal structure, described below generically as A$_3$B (see Fig. 1). The A-sublattice is connected via near-neighbor jumps between A-sites whereas B-sites are second-neighbors to each other. It is natural to suppose that diffusion of A-atoms can most readily take place by exchange with near-neighbor vacancies on the A-sublattice [4]. This is because the migration enthalpy is relatively small and because the energy of the crystal is not increased as a result of the exchange process by creation of additional point defects. Alternatively, exchange of an A-atom with neighboring B-vacancy or of a B-atom with neighboring A-vacancy would lead to creation of an antisite atom. Such an enthalpy increase is relatively large in line compounds, therefore tending to impede formation of antisite atoms. Calculations have shown that the migration enthalpy for an atom to jump to a vacant site beyond the first atomic shell is high in close-packed lattices such as this, [5] making such jumps unobservable.

For self-diffusion in a cubic structure, the diffusivity on a sublattice is given by [6]

$$D = \tfrac{1}{6} f \ell^2 w, \qquad (1)$$

in which $\ell$ is the jump-distance, $w$ is the jump-frequency of the host atom (inverse of the mean residence time), and $f$ is the correlation factor for diffusion. Measured temperature dependences of $w$ were found to satisfy Arrhenius expressions of form $w = w_0 \exp(-Q/k_B T)$ very well in all phases studied, in which $Q$ is the jump-frequency activation enthalpy and $w_0$ is a jump-frequency prefactor. For completely uncorrelated jumps, the correlation factor $f = 1$. However, a tracer atom that has just exchanged with a vacancy has a greater than average probability for making a reverse jump back into the vacancy, making $f<1$. If, after an exchange, the vacancy jumps away to the second neighbor shell of the tracer, there is still a finite probability that in subsequent jumps the vacancy will return next to the probe "on the same side", allowing for a reverse jump, again maintaining $f<1$. When the vacancy concentration is very low, $f$ can be calculated via summation methods described in [2, 3] and is "geometrical" in nature. For vacancy diffusion on the A-sublattice in the Cu$_3$Au structure, $f = 0.689$ [7]. For higher vacancy concentrations, the jumping atom may encounter a "different" vacancy whose motion is uncorrelated with that of previous jumps, increasing the averaged value of $f$ closer to 1. While other diffusion mechanisms are conceivable, for example if an A-atom makes transitory jumps to the B-sublattice, they are unlikely to be important due to the increased energy required to create an antisite defect [4].

Self-diffusion on the B-sublattice is more problematic. Direct exchange with B-vacancies in the second-neighbor shell is strongly impeded owing to the long jump-distance and steric hindrance [5]. Considering only near-neighbor jumps, a B-atom can move from a B-site by jumping into an A-vacancy, creating a B-vacancy and $B_A$ antisite atom. Next, the $B_A$-atom may jump back to the vacant B-site, or one of the A-atoms on the 11 other sites neighboring the B-vacancy can jump into it, creating a second antisite defect, $A_B$, and A-vacancy, increasing the crystal enthalpy even more. Sequences of six-jumps have been found that allow for diffusion of B-atoms without increasing the overall crystal enthalpy [8]. Finally, antisite $B_A$ atoms formed by such jumps or present in B-rich, nonstoichiometric alloys can migrate as impurities on the A-sublattice [4].

**Impurity diffusion**. Lattice locations of the solute atoms were readily observed in the present studies via nuclear quadrupole interactions. A solute might strongly prefer sublattice A or B, or to appreciably occupy both sublattices, with site-fractions on the two sublattices changing with sample composition [9, 10] and/or temperature [9]. The solute may attract or repel the vacancies that mediate diffusion, thereby modifying the concentration of vacancies in the first neighbor shell and, consequently, the diffusion rate. Finally, local enthalpy barriers to migration will in general be modified by differences in atomic sizes and chemical interactions of solute and host atoms. This allows one to write the jump-frequency $w$ of a solute atom on the A-sublattice heuristically as

$$w = w_0 \exp(-Q/k_B T) \approx w_0 \exp(-(E_F + E_B + E_M)/k_B T), \qquad (2)$$

in which an effective activation enthalpy $Q$ is expressed as the sum of an effective formation enthalpy for A-vacancies, a binding enthalpy between solute atom and vacancy, and an effective migration enthalpy. $E_F$ is a property of the host compound whereas $E_B$ (which would be zero for self-diffusion on a connected sublattice) and $E_M$ depend both on the host compound and tracer.

**Five-frequency model**. The effects embodied in Eq. 2 can be modeled in greater microscopic detail using a "five-frequency" model originally developed to describe solute diffusion in pure metals [11] but that is equally applicable for diffusion on the A-sublattice in the $L1_2$ structure. Figure 2 summarizes the model, with the five frequencies defined in the inset table. The jump-frequency measured using PAC is equal to the product of the exchange rate $\omega_2$ and concentration of vacancies in the eight A-sites neighboring the tracer atom. Explicit expressions for the correlation factor can be obtained in terms of the five frequencies [12]. Examination of the figure shows that the correlation factor will depend strongly on relative magnitudes of the frequencies, with several notable scenarios similar to those described in [12]. (1) For self-diffusion, all rates equal $\omega$, the jump-frequency of a vacancy far from the solute. (2) When the exchange rate is much less than the dissociation rate, $\omega_2 \ll \omega_3$, the vacancy's position relative to the solute atom will get randomized between exchange jumps, bringing $f$ close to 1. (3) A large "rotation" rate $\omega_1$ of vacancies to equivalent positions next to the solute can only partially reduce the correlation coefficient. (3) When $\omega_2 \gg \omega_1, \omega_3, f \cong 0$ since the solute atom and vacancy mostly "rattle" back and forth at a high rate without effective mass transport (the diffusivity is small for the given jump frequency). (4) If the vacancy and solute are tightly bound, $\omega_3 \ll \omega_1, \omega_2$, they may migrate together as a molecule. Since jump-rates are thermally activated, the correlation factor also will in general be temperature dependent, unlike the constant factor for self-diffusion. Therefore, activation enthalpies for impurity diffusion and the jump-frequency will in general differ by an "effective" activation enthalpy for the correlation factor: $Q_D - Q \cong Q_f$. Although the five-frequency model is plausible, no experimental method has been found to measure the separate frequencies.

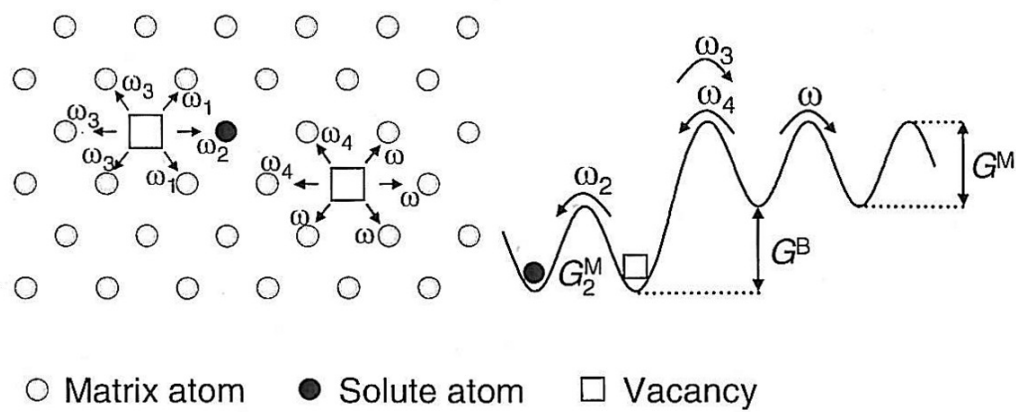

$\omega_2$: solute-vacancy exchange rate
$\omega_1$: rotation rate of the solute-vacancy pair
$\omega_3$: dissociation rate of the solute-vacancy pair
$\omega_4$: association rate of the solute-vacancy pair
$\omega$: vacancy-atom exchange rate in the solvent

○ Matrix atom   ● Solute atom   □ Vacancy

*Fig. 2.* (Left) Five-frequency model for impurity diffusion. (Right) "Energy landscape" for vacancy jumps in the neighborhood of a solute atom. (Reproduced from ref. 2, [22]).

The survey of experimental results for jump-frequencies provided below will show that jump-frequencies measured using nuclear relaxation vary by large factors even among chemically similar series of phases. One wishes to identify the enthalpy factor(s) in eq. 2 that change as one goes along a series. Some measured jump frequencies and implied diffusivities are extraordinarily large in absolute terms. For example, extrapolation of measured jump frequencies of $^{111}$Cd in indium-rich $In_3La$ [14] to room temperature and to the melting temperature (1411 K) are, respectively, 800 Hz and $10^{10}$ Hz, corresponding, respectively, to unusually high diffusivities of $2 \times 10^{-15}$ and $4 \times 10^{-6}$ cm$^2$/s using Eq. 1 (assuming $f = 1$). Possibly, correlation factors are much less than one and the solute and vacancy are only "rattling" back and forth, but one does not know.

**Experimental measurements**

Perturbed angular correlation measurements were made using decay radiations of $^{111}$In/Cd probe atoms. Experimental methods for PAC measurements are described in detail elsewhere and the methodologies are well-established [9, 10, 13]. Experiments have been carried out on ~28 phases of generic formula $A_3B$ having the $L1_2$ structure, in which A= (In, Sn, Ga, Al, Pd) and B is a rare-earth element. These include measurements on indides [14, 15, 16, 17], stannides [16, 18], gallides [19], aluminides [19], and palladides [20, 21]. For 14 indide and stannides phases, measurements were made at both opposing boundary compositions of the $L1_2$ phase field. Thus, ~40 samples were studied, each typically at 6 different temperatures, for a total of ~250 separate PAC spectra, each of which was fitted to obtain the mean jump-frequency of $^{111}$Cd tracer atoms on the A-sublattice. Measurement times needed to accumulate spectra with good statistical precision were typically 1-2 days, making for a total measurement time of a year.

Measurements were made at temperatures in the range 150-1000°C. Samples were made by melting appropriate quantities of high-purity metals together with carrier-free $^{111}$In activity under argon in a small arc-furnace. Sample masses were typically less than 100 mg to reduce non-resonant absorption of gamma-rays emitted in a cascade following the 4.0-day decay of $^{111}$In into $^{111}$Cd. The mole fraction of $^{111}$In was very low in all samples, about $10^{-10}$ to $10^{-11}$.

PAC perturbation functions were determined by accumulating time-delayed coincidence spectra between the 247 and 183 keV radiations emitted by $^{111}$Cd following decay of $^{111}$In, using four-detector spectrometers. Coincidence spectra measured at relative detector angles of 90° and 180° with respect to the sample were geometrically combined in order to obtain the perturbation function of the long-lived, 120-ns, 247-keV intermediate state. Experimental perturbation functions $G_2(t)$ (colloquially "PAC spectra") exhibit perturbations of nuclear spin orientations of the intermediate state caused by interactions between nuclear quadrupole moments of the excited state and electric-field gradients (EFG) at the nuclear sites. The EFG at a site depends on the local environment of neighboring charges, and produces a time-dependent quadrupolar perturbation characterized by a fundamental interaction frequency $\omega_Q \sim eQV_{zz}/h$, in which Q is the quadrupole moment of the nucleus and $V_{zz}$ is the principal component of the EFG tensor. A typical value of $\omega_Q$ is ~10 MHz.

Site-preferences. The EFG tensor depends on the point symmetry of the site occupied by the $^{111}$In probe parent. For B-sites, the symmetry is cubic, in which case the EFG tensor is zero, with the interaction frequency also zero. For A-sites, local symmetry leads to an axially-symmetric EFG tensor and non-zero interaction frequency. Accordingly, it is trivial to identify whether probe atoms occupy A-sites or B-sites in an L1$_2$ phase by the difference in observed signals. In general, a PAC spectrum for impurities in the L1$_2$ phase will exhibit a superposition of two signals: a zero-frequency signal, visible as a vertical offset in the time domain, and a non-zero frequency signal.

Jump-frequencies. For $^{111}$Cd tracer atoms located on A-sites at the start of the decay of the parent $^{111}$In atoms, jumps to neighboring A-sites such as shown by arrows in Fig. 1 reorient the tetragonal, principal axis of the EFG tensor by 90 degrees. The reorientation causes decoherence, or "damping", of the quadrupolar perturbation function. When the jump frequency $w$ is less than the quadrupole interaction frequency, $w << \omega_Q$, (in the so-called "slow fluctuation" regime) the dynamically relaxed perturbation function is given in very good approximation by [14]

$$G_2(t) \cong \exp(-wt) \cdot G_2^{static}(t), \qquad (3)$$

in which $G_2^{static}(t)$ is the static quadrupolar perturbation function observed in the absence of fluctuating EFGs and the multiplying factor $\exp(-wt)$ accounts for the dynamical relaxation. In the "fast fluctuation" regime, $w > \omega_Q$, the form of the perturbation function differs significantly from that given by Eq. 3, but will not be considered further here (see [14] for a fuller description). For the $^{111}$In/Cd probe, the PAC method is sensitive to jump frequencies within the range 1 MHz to 1 GHz. Jump frequencies below 1 MHz do not modify the perturbation function sufficiently to be accurately fitted due to the finite 120-ns lifetime of the PAC level. Jump-frequencies above 1 GHz lead to motional averaging of the EFGs and, for diffusion on the A-sublattice, to an average EFG of zero. Over the sensitive range of variation of $w$, one can fit experimental perturbation functions with numerical calculations of exact functions that go beyond the approximate form shown in Eq. 3 and provide highly accurate values of $w$ (see [14] and references therein). No evidence was observed of separate ensembles of probes with different jump frequencies. In addition, measured temperature dependences of jump-frequencies were always found to obey the Arrhenius form $w = w_0 \exp(-Q/k_BT)$, with no evidence of curvature that would indicate the existence of more than one jump mechanism. Taken together, these findings suggest that a single diffusion mechanism was common to all probes in each measurement.

**Results**

Indides. Since $^{111}$In is a host-element nuclide, daughter Cd tracer atoms reside on the A-sublattice at their time of creation. Fig. 3 shows representative PAC spectra measured at 1125 K for pairs of In$_3$R phases (R= Pr, Tm) having In-rich (A) and In-poor (B) boundary compositions of the L1$_2$ phase field. Differences in composition of A and B samples are unknown but probably of order several tenths of a percent from the stoichiometric composition. The spectrum for In$_3$Tm (A) shows a slightly damped static quadrupole perturbation having a period of about 100 ns attributed to $^{111}$In/Cd probe atoms on the In-sublattice. The spectrum for In$_3$Tm (B) is much more damped than In$_3$Tm (A), due to a greater jump frequency, $w \approx 10 MHz$, equal to the inverse of the damping time of ~100 ns. A similar description applies for the In$_3$Pr spectra shown on the left. Fig. 4 shows Arrhenius plots of fitted jump frequencies for A and B samples of the two phases shown in Fig. 3.

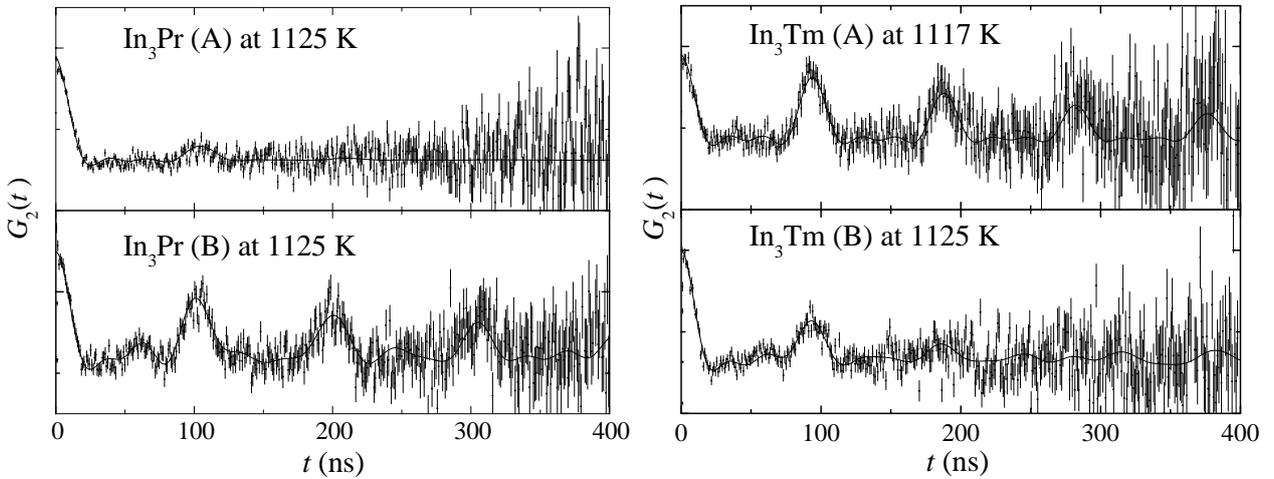

Fig. 3. PAC perturbation functions measured at about 1125 K for In-rich (A) and In-poor (B) samples of In$_3$Pr (left) and In$_3$Tm (right). The vertical scale is in arbitrary units. After [17].

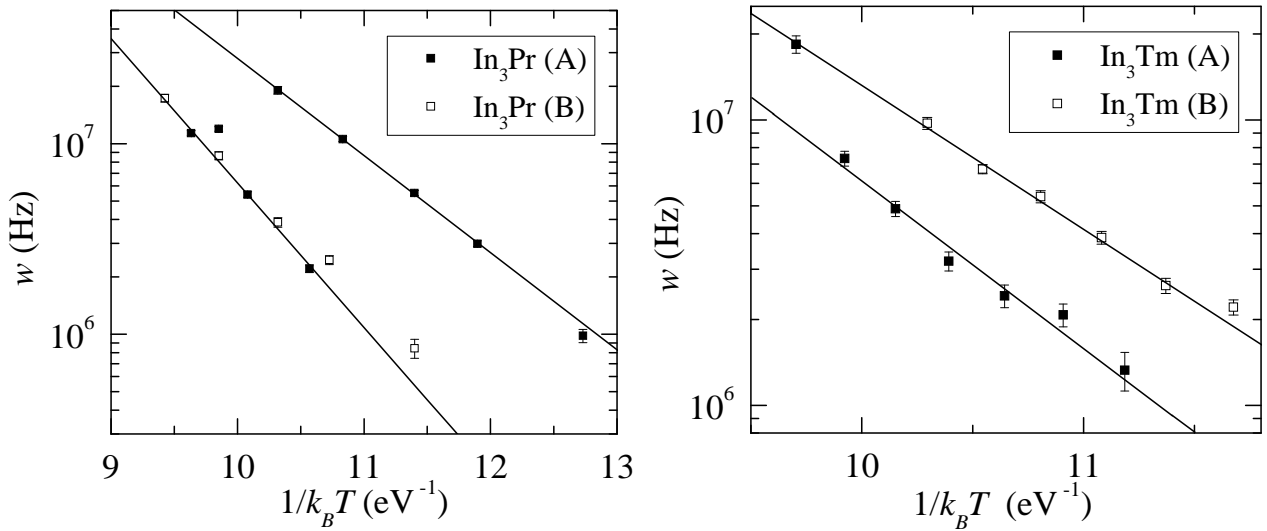

Fig. 4. Arrhenius plots of fitted jump frequencies of Cd probe atoms in In-rich (A) and In-poor (B) samples of In$_3$Pr (left) and In$_3$Tm (right). After [17]. Measured filled symbols on the lower line in the figure on the left were observed in later measurements and explained as caused by gradual evaporation of indium from the indium-rich sample during the measurements, until the average composition crossed the phase field quickly and became anchored at the In-poor boundary composition for the later measurements.

A large contrast is observed between the Pr- and Tm-phases in Fig. 4: the jump frequencies are greater for In-rich $In_3Pr$, but for In-poor $In_3Tm$. Indeed, jump frequencies were observed to be greater for all In-rich samples for the light-lanthanide elements R= La,Ce,Nd,Pr, yet greater for In-poor samples for the heavy-lanthanide elements R= Sm,Gd,Tb,Dy,Er,Tm,Lu [17].

This difference was attributed in ref. [17] to a change in diffusion mechanism along the $In_3R$ series. In general, it can be shown that as the mean composition of a compound such as $A_3B$ becomes more A-rich, the B-vacancy concentration must increase monotonically, and vice versa. Therefore, a greater concentration of A-vacancies should be present at the B-rich boundary composition, thereby increasing the overall jump frequency probability under the assumed A-sublattice vacancy diffusion mechanism. This is the observed behavior for the heavy-lanthanide indides but not the light-lanthanide indides. The suggestion from [17] was that diffusion in the light-lanthanide indides is dominated by B-vacancies, although the precise mechanism was unclear.

Stannides. Measurements were made on $Sn_3R$ phases (R= La,Ce,Pr,Nd,Sm,Gd). For Sn-poor phases, In/Cd probe atoms predominantly occupied Sn sites on the A-sublattice. For Sn-rich samples, probes only occupied Sn-sites in $Sn_3La$ and $Sn_3Ce$, and R-sites in the others. Unlike in the indide series, jump frequencies measured for opposing boundary compositions in $Sn_3La$ and $Sn_3Ce$ exhibited "normal" behavior, with greater jump-frequencies in the Sn-poor alloys, consistent with the greater expected concentration of Sn-vacancies.

Gallides and Aluminides. Probes were only observed on A-sublattices in A-poor alloys, consistent with the heuristic rule that impurities in compounds tend to occupy sites of an element in which there is a deficiency [9]. In B-poor alloys, probes occupied B-sites, also as expected.

Palladides. $Pd_3R$ phases with $L1_2$ structure exist for all lanthanide elements. Measurements were made on R= La,Ce,Pr,Nd,Sm,Eu,Tb,Er,Yb,Lu samples with both boundary compositions. For Pd-rich samples, In probes invariably occupied the R (or B) sublattice, consistent with the heuristic rule. For Pd-poor alloys with R= Lu,Yb,Er,Tb, probes strongly preferred to occupy Pd-sites, but no nuclear relaxation was detected up to 1000°C. For Pd-poor alloys with R= Eu,Sm,Nd,Pr, probes tended to occupy both Pd- and R-sites, with site-fractions changing as a function of temperature. For Pd-poor alloys with R= Eu,Sm, In-solutes preferred Pd-sites at low temperature, with site fractions of probes on R-sites increasing with temperature. For R= Nd,Pr, it was the reverse; In-solutes preferred R-sites at low temperature, with site-fractions on Pd-sites increasing with temperature. $Pd_3Ce$ gave complex spectra, and $Pd_3La$ exhibited no probe atoms on the Pr-sublattice at all for either boundary composition. Thus, the general trend in Pd-poor alloys was for decreasing stability of In-probes on the Pd-sublattice as one ran along the entire series from R=Lu to R=La.

Jump-frequencies were too small to be measured at all for temperatures up to 1000°C in R= Lu,Yb,Er,Tb alloys. Relaxation was detected for Cd-probes on the Pd-sublattice for R= Eu,Sm,Nd,Pr, with jump-frequencies increasing in the sequence R= Sm,Eu,Nd,Pr. While the result for Eu appears out of order, the jump-frequency activation enthalpies can be seen to increase monotonically when plotted versus lattice parameters of the phases (Eu is a "divalent" lanthanide and $Pd_3Eu$ has a greater lattice parameter than $Pd_3Sm$).

Overview for all systems. Figure 5 summarizes jump-frequency measurements for all series studied [20]. In the figure are plotted the inverses of the temperatures at which the jump-frequency is equal to 10 MHz is plotted versus the lattice parameters of the $L1_2$ phases. This was found to be a convenient way to characterize the jump-frequency behavior of each sample using a single parameter obtained from fits of the temperature dependences to Arrhenius forms, such as in Fig. 4. Points vertically higher in the plot have greater jump-frequencies at lower temperatures. A tabulation of fitted values of $w_0$ and $Q$ for all samples is given in [23]. Two-thirds of the values of $Q$ range between 1.0-1.5 eV, with the lowest value being 0.53(1) eV for In-rich In$_3$La. Half of the values of $w_0$ lie in the range 1-10 THz, roughly equal to the vibrational frequency of an atom, which seems reasonable for a vacancy-diffusion mechanism.

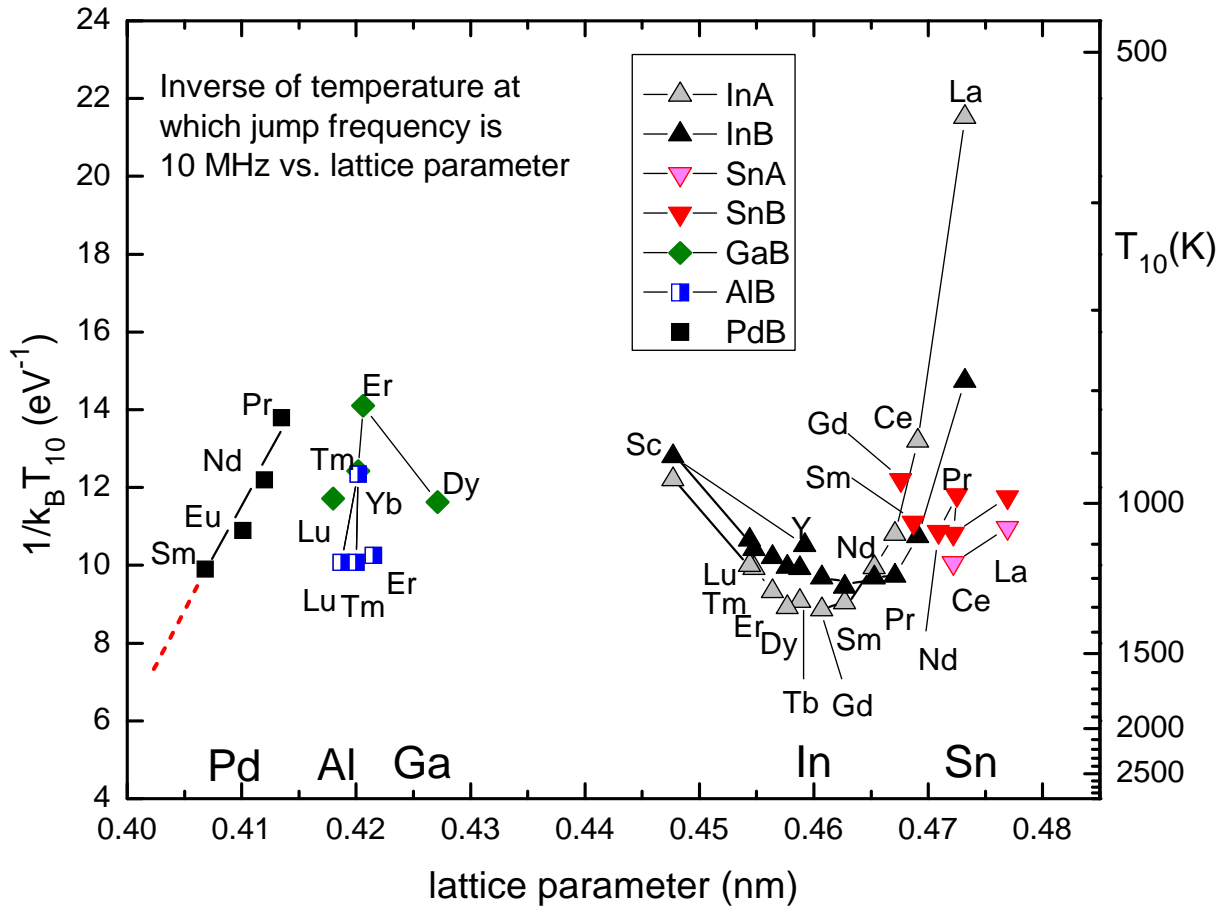

Fig. 5. Summary plot of diffusional jump-frequencies in L1$_2$ phases X$_3$R of $^{111}$Cd tracer atoms on the X-sublattice, with X= Pd, Al, Ga, In, Sn, and R= the indicated rare-earth elements. Shown are the inverses of temperatures at which the jump-frequencies were observed to be 10 MHz, plotted versus lattice parameters of the phases. Measurements were made on pairs of samples having X-rich and X-poor boundary compositions of the L1$_2$ phase field, designated in the figure respectively by A and B. For X= Pd, Ga, Al, tracer atoms strongly avoided the X-sublattice, so that no measurements could be made. After [20].

Examination of Fig. 5 shows the following features:
(a) There is no obvious dependence of the jump frequencies on lattice parameter over the five series of alloys.
(b) As noted before, there is a change in behavior along the indide series, with jump-frequencies observed to be greater for the In-rich phases with La, Ce, Pr, Nd, but for In-poor phases for the rest of the lanthanide elements (Sm to Lu). This was attributed in [17] to a change in diffusion mechanism. One expects the A-sublattice vacancy concentration to be greater for the R-rich boundary composition, so that the behavior observed for the Sm-Lu alloys is "normal" for the A-sublattice vacancy diffusion mechanism, while observations for La-Nd alloys are "anomalous".
(c) Unlike for the indides, behavior for stannides phases with La and Ce is "normal", with greater jump-frequencies observed for Sn-poor boundary compositions.

(d) Jump frequencies for the light-lanthanide indides (La,Ce,Pr,Nd) and palladides (Pr,Nd,Eu,Sm) are significantly greater than for heavier ones. This trend is discussed below.

**Discussion**

Non-observation of signals due to defects. Only single, axially symmetric, quadrupolar perturbations were observed in the spectra. Vacancies and other point defects next to PAC probe atoms would produce readily detected static quadrupole interactions [24], but the EFG due to the defect would add to the lattice EFG, leading to a perturbation that was not axially symmetric. This was not observed. This applies also for structural point defects that much accommodate deviations from the stoichiometric composition. Non-observation of signals that could be attributed to point defects is consistent with well-ordered $L1_2$ line-compounds, in contrast to previous measurements on systems with broad phase fields such as CoAl or PdIn [25]. Consequently, it is assumed that the vacancy concentrations responsible for diffusion in the present samples were low and that observed motional relaxation was caused by rapid passage of vacancies during the lifetime of the PAC level [24, 25].

Observed relaxation was previously attributed solely to diffusion on the A-sublattice. The damped quadrupole interaction signals observed were assumed only to arises from probe atoms starting on the A-sublattice and were fitted assuming that the relaxation was caused solely by jumps of the probe atom on the A-sublattice, as described in [14]. It cannot be excluded, however, that $^{111}$Cd tracer probes might also make some jumps to vacancies at B-sites, in which case a different model would be needed to describe the nuclear relaxation than that used in [14] and implicit in Eq. 3. Possible diffusion mechanisms involving B-vacancies were discussed in [17], and it was suggested that the most plausible one involved six-jump cycles of atoms, starting from an isolated B-vacancy.

Calculations of vacancy formation energies in $In_3La$ and $In_3Lu$. The much larger jump-frequencies observed for $In_3$(La,Ce,Pr…) than for $In_3$(Lu,Tm,Er…) and anomalously greater jump-frequencies for In-rich alloys than for In-poor ones in $In_3$(La,Ce,Pr…) suggest that there might be large changes in vacancy formation energies along the lanthanide series. To test this, John Bevington carried all-electron FLAPW density-function theory calculations of defect energies using the WIEN2k program [26] as part of his dissertation research [27]. In Table 1 are shown results for formation energies of In-vacancies at A- and B-sites in $In_3La$ and $In_3Lu$ [27].

Table 1. Formation energies of vacancies calculated for two $L1_2$ phases [eV].

|  | $In_3La$ | $In_3Lu$ |
|---|---|---|
| In-vacancy | 1.37 | 1.11 |
| R-vacancy (La or Lu) | 3.26 | 1.52 |

It can be seen that the 3.26 eV formation energy calculated for a R-vacancy in $In_3La$ is very high, much higher than the formation energy of an In-vacancy. Furthermore, the R-vacancy energy is even greater for $In_3La$ than for $In_3Lu$. These results emphatically rule out the possibility that the anomalous behavior observed in the indide series can be attributed to a low formation energy for R-vacancies among the light-lanthanide indides.

Tentative explanation for the anomalous jump-frequency behavior. A strong correlation has been noted as described above and in refs. [20, 21] between the site preferences of In parent-probe atoms and jump frequencies of Cd-daughter probes in the palladides. Proceeding along the palladide series of Pd-poor alloys from Lu, Yb, Er, Tb, Sm, Eu, Nd, Pr, La the site preference of the In-parent probe changes from strongly preferring Pd-sites (Lu, Yb, Er, Tb) to weakly preferring Pd-sites (Sm, Eu), to weakly preferring R-sites (Nd, Pr), to strongly preferring R-sites (La). At the same time, jump frequencies of Cd daughter probes on the Pd-sublattice are undetectably low for (Lu, Yb, Er, Tb), and increase in the series (Sm, Eu, Nd, Pr). While the behaviors apply to different probes (In for site preferences; Cd for jump-frequencies), it is reasonable to suppose that the change in site preference of Cd along the series mirrors that of In. Thus, the site preference for daughter Cd-probes may be changing along the series, becoming increasingly unstable for lanthanides near the lanthanum end of the series (Sm, Eu, Nd, Pr). Indeed, it is possible that the Cd-probe has changes in site-preferences that give it stability at R-sites that In-probes do not have.

In the indides, In-probes of course can only be found at In-sites. But Cd-daughter probes have an increasing preference to occupy R-sites for alloys in the sequence (Sm, Nd, Pr, Ce, La). Although a precise jump mechanism is unknown, the increasing site preference will tend to be mirrored in more rapid transfer of Cd-probes from In-sites on which they are born to R-sites. The greater magnitude of the relaxation observed for In-rich samples may reflect a very small, but appreciable concentration of R-vacancies, when one keeps in mind that the mole fraction of probes is only of order $10^{-11}$. Perhaps Cd-probes make jumps to R-sites and are then irreversibly "trapped" by a strong site preference. Possibly, this non-equilibrium jump-mechanism may lead to greater nuclear relaxation per jump than for the A-sublattice vacancy diffusion mechanism, explaining the large magnitude of the relaxation in the In-rich and Pd-rich light-lanthanide indides and palladides.

<u>Experimental determination of the correlation factor for impurity diffusion</u>. There are few methods for measuring jump-frequencies of atoms in solids and there has never been, to our knowledge, a determination of the correlation factor or its temperature dependence. The jump-frequency measurements made in this laboratory could be combined with diffusivity measurements made on the same systems, for example using indium alloys and Cd-tracers, to obtain the correlation factor via Eq. 1.

This work was supported in part by the National Science Foundation under grant DMR 09-04096 and by the Praveen Sinha Fund for Scientific Research.